\documentclass[aps,prd,preprint,superscriptaddress,showpacs]{revtex4-1}
\usepackage{graphicx}
\usepackage{dcolumn}
\usepackage{bm}
\begin{document}


\title{On the Galactic origin of ultra-high energy cosmic rays}

\author{V.N.Zirakashvili}
\affiliation{Pushkov Institute of Terrestrial Magnetism, Ionosphere and Radiowave
Propagation,\\ 108840 Moscow Troitsk, Russia }
\author{V.S.Ptuskin}
\affiliation{Institute for Physical Science and Technology, University of Maryland,\\College 
Park, MD 20742, USA}
\author{S.I.Rogovaya}
\affiliation{Pushkov Institute of Terrestrial Magnetism, Ionosphere and Radiowave
Propagation,\\ 108840 Moscow Troitsk, Russia }



\date{\today}

\begin{abstract}
It is shown that the acceleration of particles by a
powerful relativistic jet associated with the activity of a supermassive
black hole in the Galactic center several million years ago may
explain the observed cosmic ray spectrum at energies higher than
$10^{15}$ eV. The accelerated particles are efficiently confined in
the extended magnetized gas halo created by the supernova and
central black hole activity just after the Galaxy formation. We found
that both the heavy and light chemical composition of ultra-high
energy cosmic rays can be consistent with observations.
\end{abstract}

\maketitle


\section{Introduction}

The prevailing point of view is that the origin of observed ultra-high energy cosmic rays (UHECRs) is extra-galactic.
It is primarily due to the observations of astronomical objects with high energetics, which is required for
accelerating to ultra-high energies, in other galaxies. These are jets in active galactic nuclei (AGN),
gamma-ray bursts, tidal disruption events, etc.
\cite{bykov12, rieger22, globus23}.

However, such objects are at times present in our Galaxy.  In particular Fermi and e-Rosita
bubbles \cite{su10,predehl20} are probably linked with the past activity of a supermassive black hole
(SMBH) in the Galactic center. Cosmological simulations of the Milky Way, and
Andromeda-like galaxies \cite{pillepich21} demonstrated a periodic activity of SMBH every $10^8$ years with
a peak mechanical luminosity of about $10^{44}$ erg s$^{-1}$.

If so, some amount of high-energy
cosmic rays may have been produced during the periods of activity. Models of this kind have
already been suggested in the past \cite{fan51,burbidge63,kulikov69,wayland72,ptuskin81, giler83,
istomin14, fujita17}. The possibility to observe these cosmic rays
critically depends on the confinement of particles in the Galaxy.

It appears that this
confinement has the potential to be better than previously thought. It is known now that the
Milky Way and other galaxies are surrounded by huge halos of hot gas \cite{bogdan22}. The gas
contains both primordial accreting gas and galactic gas that was ejected from the galaxy
during early epochs of enhanced star formation and SMBH activity. Since the galactic
magnetic fields were also ejected by the outflows we expect rather effective confinement of
particles in such extended (several hundred kpc in size) halos.

Our preliminary model of the acceleration and propagation of UHECRs from nearby SMBHs
in the Galactic center and Andromeda galaxy \cite{zirakashvili23} (Paper I) is further elaborated in the
present paper. Three components of particles accelerated in the jet were considered in this
model. The lowest energy particles are accelerated at the bow shock of the jet by the
diffusive shock acceleration (DSA) mechanism \cite{krymsky77,bell78,axford77,blandford78}.
The highest energy particles are
accelerated in the jet itself via the shear acceleration \cite{berezhko81, earl88}
 or via DSA at the termination shock of the jet.
For spectral continuity, a third
intermediate component of accelerated particles was introduced. It could be related to the
acceleration in the turbulent jet cocoon or the acceleration in the SMBH magnetosphere \cite{istomin09,
banados09, jacobson10, wei10}.

In the present paper, we concentrate on the propagation of UHECRs from the Galactic center
and check whether it could considerably contribute to the observed spectrum of UHECRs.

The paper is organized as follows.  In the next Section 2, we briefly remind our model \cite{zirakashvili23}.
Section 3 provides a description of magnetic fields in the Galactic halo. Section 4 presents
the numerical results for the propagation of particles from the Galactic center. Sections 5 and
6 contain the discussion of results and conclusions. The Appendix describes the numerical
modeling of the extended gaseous Galactic halo.

\section{Model of cosmic ray acceleration and propagation}

\begin{table}[tbp]
\centering
\begin{tabular}{|c|c|c|c|c|p{4.2cm}|}
\hline  component&$\gamma $&$\epsilon _{\max }$  & $L_{\mathrm{cr}}(E>1\ \mathrm{GeV})$& $E_{\mathrm{cr}}(E>1\ \mathrm{GeV}) $&
$k(A)/k_{\odot }(A)$\\
\hline   jet     &  1.0    &$4\times 10^{19}$ eV       &$1.3\times 10^{37}$ erg s$^{-1}$&$4\times 10^{52}$erg &
$4, A=4,\ 2(A/Z)^2, A>4$\\
\hline  bow shock &  2.2    &$6\times 10^{15}$ eV&${6.9\times }10^{39}$ erg s$^{-1}$&$2.2\times 10^{55}$erg &
$2, A=4,\ A/4, A>16$, $2A/Z,\ 4<A\le 16$\\
\hline  inner jet   &  2.2    &$2\times 10^{18}$ eV&${2.1\times }10^{39}$ erg s$^{-1}$&$6.6\times 10^{54}$erg&$0, \ A>1$\\
\hline
\end{tabular}
\caption{Parameters of the source components in the Galactic center for the model "light"}
\end{table}

\begin{table}[tbp]
\centering
\begin{tabular}{|c|c|c|c|c|p{4.2cm}|}
\hline  component&$\gamma $&$\epsilon _{\max }$  & $L_{\mathrm{cr}}(E>1\ \mathrm{GeV})$& $E_{\mathrm{cr}}(E>1\ \mathrm{GeV})$ &
$k(A)/k_{\odot }(A)$\\
\hline   jet     &  1.0    &$5\times 10^{18}$ eV       &$6.2\times 10^{37}$ erg s$^{-1}$&$2.0\times 10^{53}$erg&
$4, \ A=4,\ 80, \ A>4$\\
\hline  bow shock &  2.0    &$4\times 10^{15}$ eV&${1.1\times }10^{40}$ erg s$^{-1}$ &$3.5\times 10^{55}$erg&
$2, A=4,\ A/4, A>16$, $2A/Z,\ 4<A\le 16$\\
\hline  inner jet   &  2.0    &$1.3\times 10^{18}$ eV&${1.9\times }10^{39}$ erg s$^{-1}$&$6\times 10^{54}$erg&$0, \ A>1$\\
\hline
\end{tabular}
\caption{Parameters of the source components in the Galactic center for the model "heavy"}
\end{table}

A detailed description of our model can be found in Paper I. The calculations of cosmic ray
propagation include the spatial diffusion, energy losses, and nuclei fragmentation of protons
and nuclei traveling from the central instantaneous point source. The source produces three
components of accelerated particles. Each component has a spectrum that is described by the
equation

\begin{equation}
   q(\epsilon ,A)\propto k(A){\epsilon }^{-\gamma }
\exp {\left( -\frac {A\epsilon }{Z\epsilon _{\max }}\right) }
\end{equation}
where $\epsilon $ is the energy per nucleon, $A$ and $Z$ are the atomic mass and charge numbers respectively,
the function $k(A)$ describes the source
chemical composition and can be written in terms
of the solar composition $k_{\odot }(A)$.

The parameters of the source spectra are given in Tables 1 and 2,
as explained below.

Figure 1 shows the schematic view of the jet. The jet drives a bow shock in the interstellar
medium. This shock accelerates particles through the DSA mechanism. The chemical
composition of accelerated particles depends on the chemical composition of the interstellar
medium and the enrichment due to the preferential injection of ions in comparison to
protons. It was found in the hybrid modeling of collisionless shocks \cite{caprioli17} that this enrichment
is proportional to the ratio of the atomic mass to the charge of injected ions. Thus we
expect that function $k(A)=2k_{\odot }(A)$ for fully ionized He ions. When it comes to heavier
ions, we take into account an enhanced metallicity 2 of the Galactic bulge \cite{revnivtsev04} and assume
that ions are strongly ionized up to the charge number 8 by a powerful X-ray radiation from
the accretion disk. This gives $k(A)=Ak_{\odot}(A)/4$ for ions heavier than Oxygen and $k(A)=4k_{\odot }(A)$
for lighter ions. In other words, the ions of the CNO group are fully ionized.

The highest energy particles are accelerated in the jet itself. The most probable mechanism is
the shear acceleration that occurs in the vicinity of the boundary between the jet and the
surrounding medium \cite{seo21}. We assume that the injected ions are fully ionized in the jet.

To maintain spectral continuity, a third intermediate component with light composition is
necessary. In the present paper, we consider a pure proton third component. The existence of
this component is validated by observations. It is known that particles are accelerated in the
vicinity of the SMBH and we observe a variable in time radio, X-ray, and gamma emission in
jets \cite{blandford19}. Although this emission is probably of leptonic origin, the acceleration of
protons and nuclei is also highly probable. The association of observed astrophysical
neutrino events with blazars supports this scenario \cite{troitsky21}. An important point is that
accelerated nuclei are fully photo-disintegrated and protons are subject to energy losses in the
strong radiation field. They escape the jet via the neutron production mechanism \cite{berezinsky78}.
Because neutrons don't interact with magnetic fields, the spectrum of escaped neutrons is
similar to the proton spectrum inside the jet. Later the neutrons decay and turn into protons.
As for the acceleration mechanism, it can be either shear acceleration or acceleration during
multiple magnetic reconnection events \cite{alves18}. The latter seems to be possible close to SMBH
where the jet is Pointing dominated.

\begin{figure}
\begin{center}
\includegraphics[width=8.0cm]{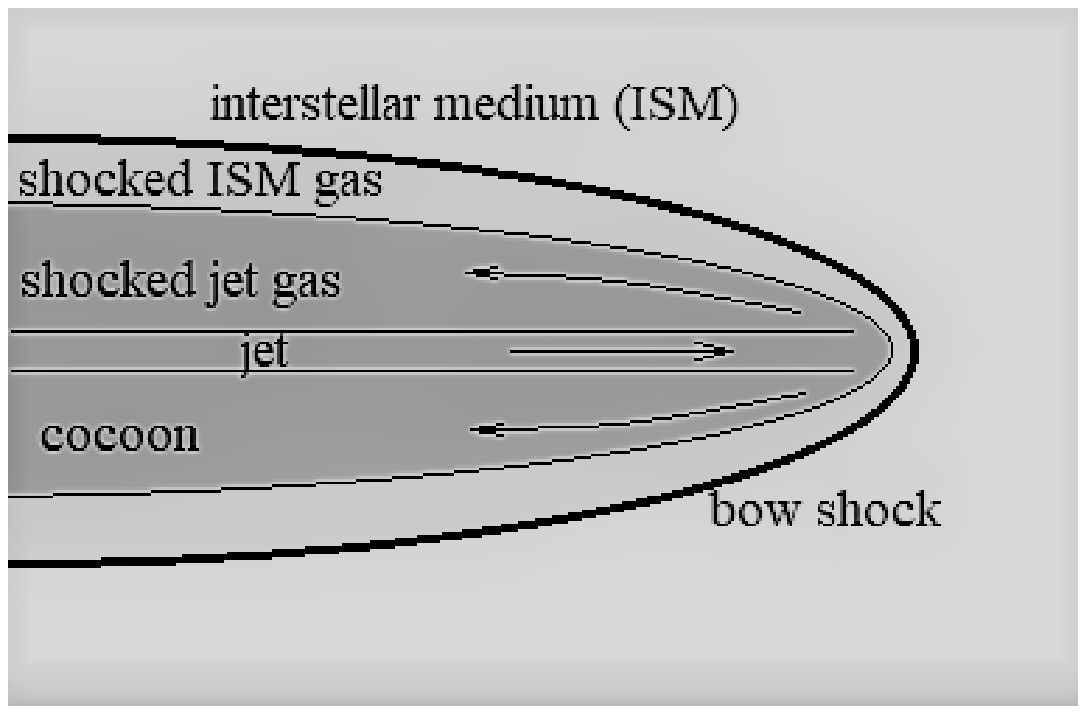}
\hfill
\includegraphics[width=7.0cm]{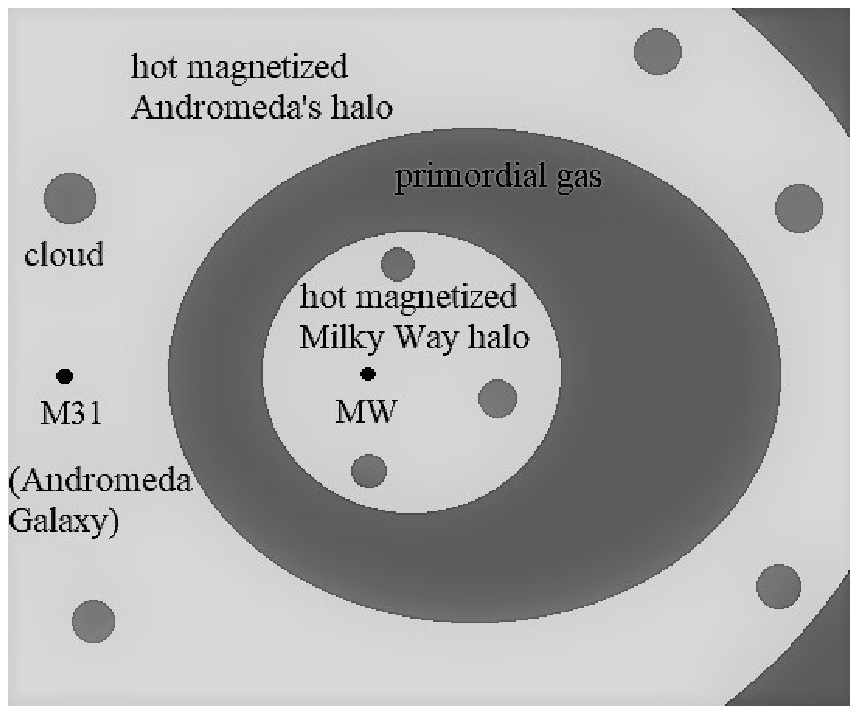}
\end{center}
\caption{ Schematic view of the jet (left panel) and hot magnetized
halos of the Milky Way (MW) and Andromeda galaxy (M31) (right panel). }
\end{figure}

The shear acceleration can reaccelerate the low-energy protons and nuclei of the intermediate
component at large distances from the SMBH when the radiation field is not strong. In this
regard, there is a correlation between the intermediate and the highest energy components.
There is an additional enrichment of nuclei by a factor of $(A/Z)^{\Delta \gamma }$ related to this
reacceleration. Here $\Delta \gamma \approx 1$ is the difference of spectral indexes.
This gives $k(A)=4k_{\odot }(A)$
for fully ionized Helium ions and $k(A)=2(A/Z)^2k_{\odot }(A)$ for fully ionized heavier ions of
the highest energy jet component. The additional factor 2 comes from the higher metallicity
of the Galactic bulge.

To describe particle diffusion, we use an analytical approximation of the diffusion coefficient
in an isotropic random magnetic field with the Kolmogorov spectrum \cite{harari14}.

\begin{equation}
D= \frac {cl_c}{3}\left( 4\frac{E^2}{E^2_c}+0.9\frac{E}{E_c}
+0.23\frac{E^{1/3}}{E^{1/3}_c}\right) , \ E_c=ZeBl_c=0.9\ \mathrm{EeV}\ ZB_{\mu \mathrm{G}}l_{c,\mathrm{kpc}},
\end{equation}
where $E$ is the energy of the particle, $B$ is the magnetic field strength and $l_c$
is the correlation length of the magnetic field. At large energies $E\gg E_c$ the scattering of particles occurs on
 the magnetic inhomogeneities with scales smaller than the particle gyroradius and
the diffusion coefficient is proportional to $E^2$. At lower energies $E\ll E_c$ the resonant scattering results in
 the energy dependence of diffusion $\sim E^{1/3}$.

\section{Hot magnetized galactic halos}

Rapid energy release and the creation of strong outflows (galactic winds) are the results of an
enhanced star formation and accretion onto the central supermassive black hole shortly after
a galaxy formation. A developing cavity was created in the circumgalactic medium by a hot gas
that was ejected and heated by a wind termination shock. In cosmological models, galaxies
with strong AGN feedback are shown to have extended "bubbles" of multi-Mpc size,
whereas galaxies with a weaker supernova feedback have smaller bubbles of sub-Mpc size \cite{aramburogarcia21}.
Figure 1 depicts a schematic view of the halos that are thought to have evolved around the
Milky Way and Andromeda galaxy.

Galactic magnetic fields are also ejected by the outflows \cite{voelk00}. It is expected that they are
amplified by a so-called Cranfill effect \cite{axford72,cranfill74} downstream of the termination shock.
Although the gas energy density is higher than the magnetic energy density just downstream
of the termination shock, the radial contraction in the incompressible expansion flow results
in the amplification of the non-radial components of the magnetic field. The magnetic field
strength increases proportional to the distance. As a result, the thermal and magnetic energies
are comparable at large distances, see Appendix. We expect isotropy of the random fields
because of the turbulent gas motions generated by clouds of accreting colder and denser
circumgalactic gas inside the bubble of the shocked galactic gas.

We can obtain a rough estimate of the magnetic field strength in the Milky Way extended
halo. It is assumed that 1 $\% $ of the Galactic $10^{11}$ stars ended their lives as supernovae. For the
standard supernova energy $10^{51}$ erg, we obtain the total energy of $10^{60}$ erg. In addition, our
SMBH in the Galactic center has the mass $4\times 10^6M_{\odot}$ and the corresponding rest mass
energy $7\times 10^{60}$ erg. Assuming that 10 $\% $ of this energy goes into the outflows during SMBH
growth and taking into account 30 $\% $ of the supernova energy we obtain a total of $10^{60}$ erg. Then
the total energy density of the gas and magnetic field is equal to $1.3\times 10^{-13}$ erg cm$^{-3}$ for the
bubble radius $R = 400$ kpc. The corresponding equipartition magnetic field strength is $1.3\mu $G.

Note that the mean magnetic field strength of $0.5\mu $G along the line of sight at 100 kpc
galactocentric distances was estimated from the recent measurements of the Faraday rotation
performed for different samples of galaxies \cite{heesen23,bockmann23}. The actual value can be higher because
of the field reversals and a lower gas number density than assumed $n=10^{-4}$ cm$^{-3}$. Hence our
rough estimate is in accordance with observations.

\begin{figure}[tbp]
\begin{center}
\includegraphics[width=7.0cm]{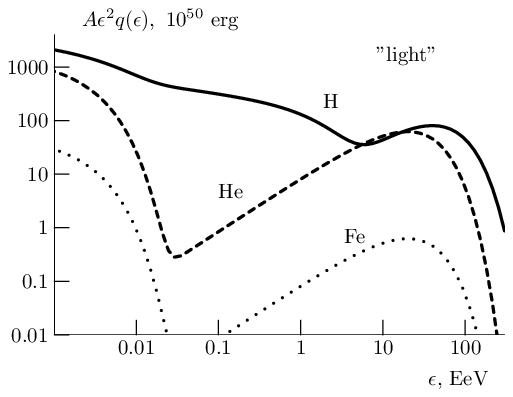}
\hfill
\includegraphics[width=7.0cm]{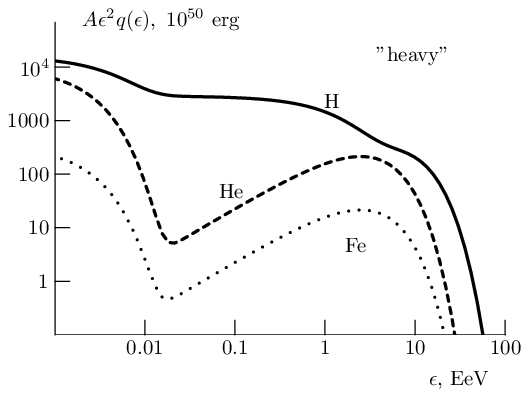}
\end{center}
\caption{ Source spectra of protons (solid line), He nuclei (dashed line) and Iron (dotted line) produced
in  the Galactic center in models "light" (left panel) and "heavy" (right panel).}
\end{figure}

\section{Numerical results}

We model the propagation of particles in the spherical simulation domain with radius
 $R=400$ kpc where an absorbing boundary condition is set. It is assumed that the Galactic center
source is in active phase every 100 million years. The age $T$ of the Fermi and e-Rosita
bubbles considered as a result of the last active phase is not exactly known. We analyze two
models of bubble formation at $T = 3$ ("light") and $T = 15$ ("heavy") million years ago. The
parameters of the source spectrum are adjusted to reproduce observations and are given in
Tables 1 and 2. They contain the cosmic ray energy of every energetic component $E_{\mathrm{cr}}$ per
one activity event and the mean cosmic ray luminosity $L_{\mathrm{cr}}$ averaged over 100 million years.
The contribution of the Galactic center in observed UHECRs is
dominated by the last active event while more ancient events are important for extragalactic contribution,
see below.

The source spectra of protons and nuclei are shown in Figure 2.

For the model "light" the enrichment of the highest energy jet
component by heavy nuclei is not needed because Helium nuclei have
no time for the photodisintegration. The random magnetic field
strength $B=1\mu $G and the correlation length $l_c=80$ kpc are
accepted. For the model "heavy" we use lower values of the magnetic
field strength $B=0.5\mu $G  and the correlation length  $l_c=40$
kpc. The heavy nuclei in the model "heavy" are 10 times more
abundant
 in comparison with the model "light". This is because these nuclei contain only 1 $\% $ of mass
in the interstellar medium
and this is not enough to explain the chemical composition of observed UHECRs.

For such magnetic field and correlation length the scattering
free path of particles $\lambda $ is small enough to justify the use
of the diffusion approximation. For example it is close to $\lambda
=130$ kpc for highest energy Helium nuclei with energy $E=7\times
10^{19}$ eV in the model "light".

We also calculate a possible extragalactic contribution for both models. We use in this case
the simulation domain with a radius $R=2.4$ Mpc and a reflecting boundary condition that is
a zero gradient of cosmic ray distribution at the boundary. This implies the mean distance
4.8 Mpc between extragalactic sources and corresponds to the source number density of 0.01 Mpc$^{-3}$.
All sources have identical spectra shown in Figure 2. The particles are released every 100 Myrs.
The magnetic field strength $B=10^{-10} $G was assumed in this case.

The calculation is performed up to the maximum redshift $z=1$
 in a flat universe with the matter density
$\Omega_{m}=0.3$, the dark energy density $\Omega_{\Lambda}=0.7$,
 and the Hubble parameter  $H=70$ km s$^{-1}$ Mpc$^{-1}$ at the current epoch.
The strong evolution of sources with a factor $(1+z)^4$ is taken into account.

For calculations of distribution for atmospheric depth of shower maximum $X_{max}$ we
use the analytical parametrization from \cite{dedomenico13,PAO23}.

The results are shown in Figures 3-7.

The spectra observed at the solar system location at the galactocentric distance $r=R_{\odot }=8.5$ kpc
 are shown in Figure 3. Both models reproduce the observed all-particle spectrum.
The "light" model is in better agreement with the observed chemical
composition if the interaction model QGSJetII-04 is used (see Figures
4 and 5). The unusual bump at the variance $\sigma (X_{max})$ curves at
the energy $10^{17}$ eV appears because protons of the intermediate component and Iron nuclei of
the bow shock component give the main input in all-particle spectrum  at these energies.
The Iron nuclei and protons have large difference of the mean depth $\langle X_{max}\rangle $ and this result
 in the large variance $\sigma (X_{max})$.
The model "heavy" is preferable for the explanation of the observed
anisotropy (see Figure 7). Relatively high anisotropy at PeV
energies in the model "light" is not a serious problem because the
anisotropy at these energies is strongly influenced by the local
Galactic magnetic fields. This effect is not taken into account in
the present study. The numerical value of the calculated anisotropy
$\delta $ is close to $\delta =1.5R_{\odot }/cT$ which is the
anisotropy of the instantaneous point source in the infinite space.
Its value can be higher if the Galaxy is shifted from the halo
center (see the right panel of Figure 7 and the discussion
below).

\begin{figure}[tbp]
\begin{center}
\includegraphics[width=7.0cm]{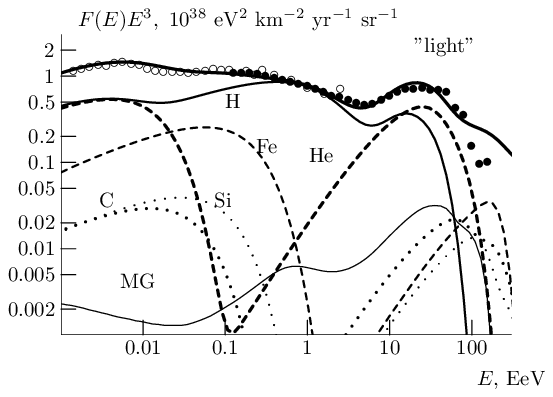}
\hfill
\includegraphics[width=7.0cm]{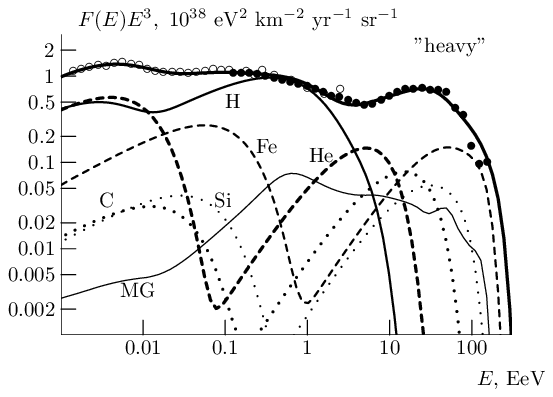}
\end{center}
\caption{ Spectra of different elements and all-particle spectrum
(thick solid line) produced in  Galactic center and observed at the
Earth position in models "light" (left panel) and "heavy" (right
panel). A possible metagalactic contribution in the all particle
spectrum (MG) is shown by the thin solid line. Spectra of Tunka-25,
Tunka-133 array (\cite{budnev20}, open circles) and PAO
(\cite{PAO21}, energy shift +10$\% $, black circles) are also
shown.}
\end{figure}

\begin{figure}
\begin{center}
\includegraphics[width=7.0cm]{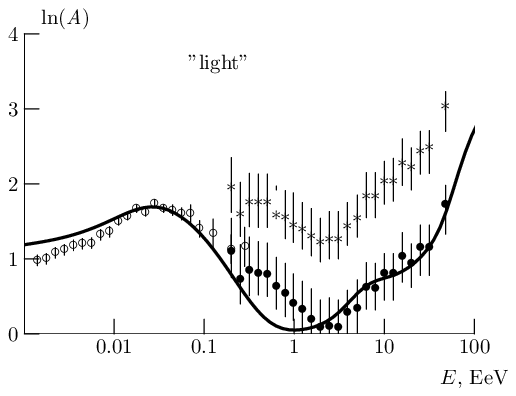}
\includegraphics[width=7.0cm]{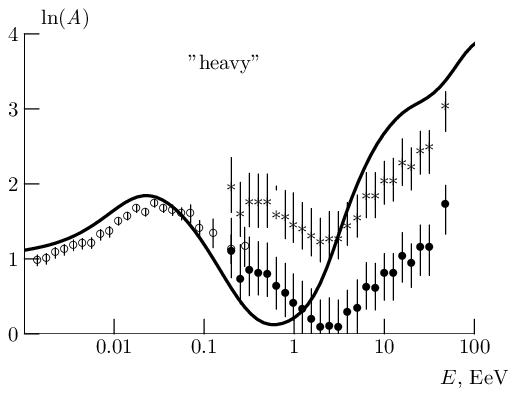}
\end{center}
\caption{ Calculated mean logarithm of atomic number A (solid line) for the model "light" (left panel) and
model "heavy" (right panel). The
measurements of Tunka-133, TAIGA-HiSCORE
 array (\cite{prosin22} open circles) and PAO
(hadronic interaction model QGSJetII-04 (black circles) and SIBYLL2.3 (asterisks),
energy shift +10$\% $ \cite{bellido17}) are also shown.}
\end{figure}

\begin{figure}
\begin{center}
\includegraphics[width=7.0cm]{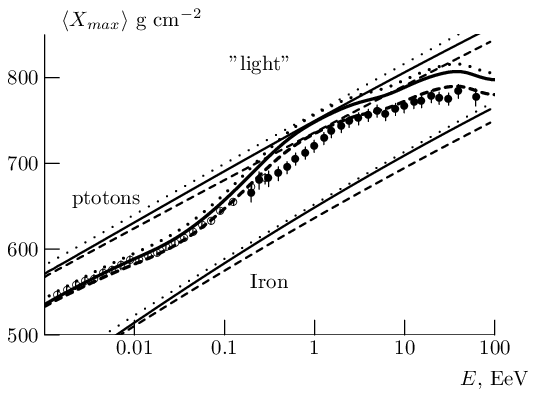}
\includegraphics[width=7.0cm]{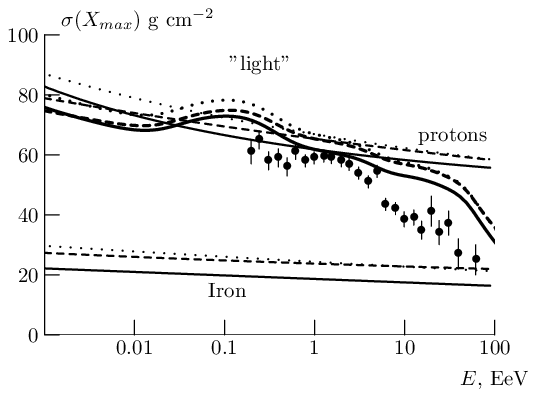}
\end{center}
\caption{ Calculated mean atmospheric depth of shower maximum $\langle X_{max}\rangle $ (left panel, thick lines),
its variance $\sigma (X_{max})$
(right panel, thick lines) for the model "light" and
the corresponding curves for pure proton and Iron composition (thin lines) .
The hadronic interaction models used are EPOS-LHC (solid lines), QGSJetII-04 (dashed lines)
and SIBYLL2.3d (dotted lines).
The measurements of Tunka-133, TAIGA-HiSCORE
 array (\cite{prosin22} open circles) and PAO (energy shift +10$\% $ \cite{yushkov19}, black circles)
are also shown. }
\end{figure}

\begin{figure}
\begin{center}
\includegraphics[width=7.0cm]{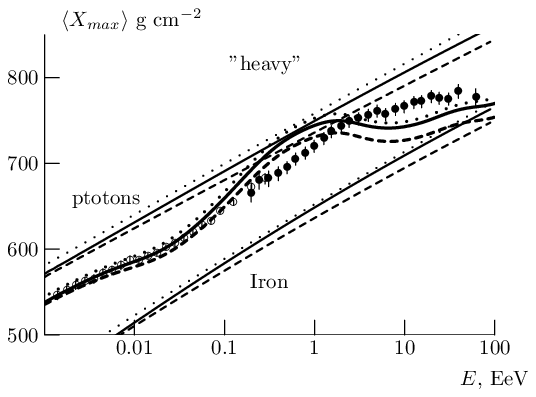}
\includegraphics[width=7.0cm]{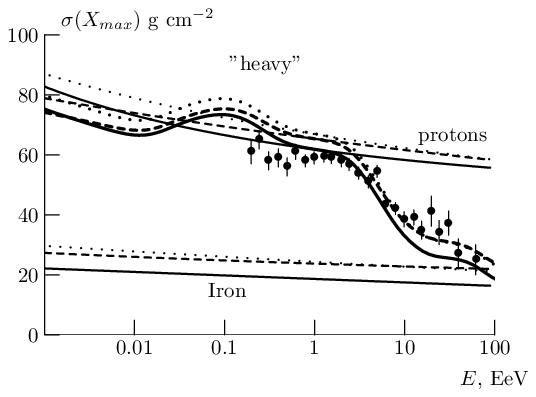}
\end{center}
\caption{ Similar to Figure 5, but for model "heavy". }
\end{figure}

\begin{figure}
\begin{center}
\includegraphics[width=7.0cm]{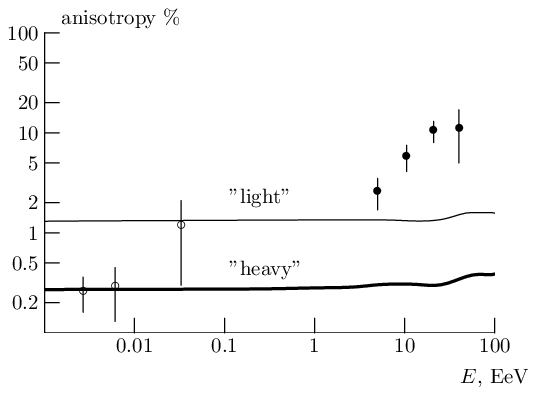}
\includegraphics[width=7.0cm]{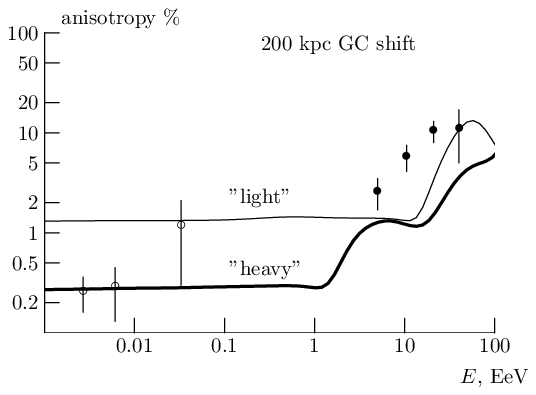}
\end{center}
\caption{ Calculated cosmic ray anisotropy (solid lines) for the Milky Way situated
in the center of the extended halo (left panel) and shifted on 200 kpc 
in the Andromeda direction (right panel).
The results
of PAO (energy shift +10$\% $, \cite{aab18} black circles) and the
KASCADE-Grande experiment (\cite{chiavassa15} open circles) are also
shown.}
\end{figure}

\section{Discussion}

The maximum energy of particles  accelerated at the nonrelativistic
bow shock is determined by the nonresonant cosmic ray streaming
instability \cite{bell04} (see Paper I for details)

\begin{equation}
\epsilon ^b_{\max }=\frac {\eta _{\mathrm{esc}}}{2\ln (B/B_b)}e\sqrt{\beta _{\mathrm{head}}L_{\mathrm{j}}c^{-1}}
=1.73\times {10^{19}} \mathrm{eV} \ \frac {\eta _{\mathrm{esc}}}{2\ln (B/B_b)} \beta ^{1/2}_{\mathrm{head}} \left(
\frac {L_\mathrm{j}}{10^{44}\mathrm{erg}\ \mathrm{s}^{-1}} \right)
^{1/2}
\end{equation}
Here $\beta _{\mathrm{head}}$ is the ratio of the speed of bow shock
"head" to the speed of light $c$,
 $L_{\mathrm{j}}$ is the total power of two opposite directed jets, $\eta _{\mathrm{esc}}$ is the ratio of the energy
 flux of runaway accelerated particles to the kinetic flux  of the shock.
The logarithmic factor in the denominator corresponds to the situation when the seed magnetic field $B_b$ is
amplified in the
 upstream region of the shock up to values of $B$ via cosmic ray streaming instability.

The parameter $\eta _{\mathrm{esc}}$ is close to 0.01 for shocks where the pressure of accelerated particles is
of the order of 0.1 of the shock ram pressure and can be higher at cosmic-ray modified
shocks. The protons can be accelerated up to multi PeV energies at the jet bow shocks
at $\eta _{\mathrm{esc}}=0.01$, $\beta _{\mathrm{head}}=0.1$ and $\ln (B/B_b)=5$.

Our modeling shows that particles with energies above $10^{15}$ eV can be produced in the
Galactic center and observed at the Earth.

Below PeV energies, the particles have no time to reach the Earth
and we expect a smooth low energy cut-off of the spectrum. Lower energy
particles are probably produced in Galactic supernova remnants. In
this regard, our scenario is similar to the model with a nearby
source \cite{erlykin97}. This model was suggested for the
explanation of the "knee" in the observed cosmic ray spectrum.
The similar in spirit origin of UHECRs diffusing
 from the point source in the Galactic center was also considered in the past \cite{kulikov69}.

It is known that the electric potential difference is a reasonable
estimate for the maximum energy of particles accelerated at
quasi-perpendicular shocks \citep{zirakashvili18}. For example,
single-charged anomalous cosmic rays are accelerated up to hundreds
MeV at the solar wind termination shock with the electric potential
200 MV {\cite{cummings87}}. The jet electric potential is also a
good estimate for the maximum energy as seen in trajectory
calculations \cite{alves18, mbarek19}.

The corresponding value is

\begin{equation}
\epsilon ^j_{\max }=e\sqrt{\beta _{\mathrm{j}}L_{\mathrm{mag}}c^{-1}}
=1.73\times {10^{19}} \mathrm{eV} \ \beta _{\mathrm{j}}^{1/2} \left(
\frac {L_\mathrm{mag}}{10^{44}\mathrm{erg}\ \mathrm{s}^{-1}} \right)
^{1/2}
\end{equation}
where $L_{\mathrm{mag}}$ is the magnetic luminosity of two opposite jets,
see Paper I for details.

So the jet power $\sim 10^{45}$ erg s$^{-1}$ is needed to achieve the maximum energy of $4\times 10^{19}$ eV in
the "light" model. It means that the Galactic center SMBH with Eddington luminosity
$L_{\mathrm{Edd}}=5\times 10^{44}$ erg s$^{-1}$ was an Eddington or super-Eddington source
during the past active phase.
The duration of this phase was only 30 kyrs to supply $10^{57}$ erg of energy in e-Rosita bubbles.

A similar jet power is needed in the model of e-Rosita and Fermi bubbles formation \cite{yang22}. In
this model, a short energetic event in the Galactic center 2.6 million years ago produced jets
moving in the Galactic halo. After the jet's disappearance, the bow shock of the jet
propagated to larger heights and is observed now as the e-Rosita bubbles. The Fermi bubble
is a heated jet material inside the e-Rosita bubbles. The main difficulty of the model is a high
shock speed of 1-2 thousand km s$^{-1}$. The lower shock speed $\sim 350$ km s$^{-1}$ was inferred
from the gas temperature in e-Rosita bubbles \cite{predehl20}. However, recent X-ray observations show
the presence of more hot gas \cite{gupta23} and the shock speed can be higher.

It was also found that the properties of young stars in the vicinity of Sagittarius A can be
explained if they were formed from the massive gas shell ejected by a central energetic
super-Eddington outflow 6 million years ago \cite{nayakshin18}.
A similar high ionization energetic event 3.5 million years ago is needed for the explanation
of the ionization cones in the Galactic halo \cite{blandhawthorn19}. Ionization cones of this kind
 produced by
AGN outburst 65 kyrs ago also exist in the vicinity of Seyfert galaxy NGC 5252\cite{wang24}.

The last Galactic center activity was probably distributed in time. For example, it began
15 million years ago with a moderate power and produced the shock with the present speed of
350 km s$^{-1}$. There was an additional powerful energy release at the end of activity 3 million
years ago that produced the Fermi bubbles. This age of the Fermi bubbles is also in
agreement with the X-ray absorption study \cite{miller16}.

In this regard the past activity in the Galactic center is similar to the activity of Narrow Line Seyfert 1
 (NLSy 1) galaxies. This
 Seyfert-like activity with the Eddington luminosity is observed in spiral  galaxies with small or moderate SMBHs.
About 7 percent of these galaxies have jets. The jets directed to us are similar
to blazars and are observed as powerful gamma ray sources \cite{d'ammando19}.
The number of all jetted NLSy 1 galaxies
corresponds to the fraction $10^{-4}-10^{-3}$ of the bright galaxies. This gives an estimate for
the duration of an active
phase $10^4-10^5$ years similar to the  parameters of the model "light".

On the other hand, the lower maximum energy in the model "heavy" can be achieved with the jet
power of the order of several percent of the Eddington luminosity. The duration of the active
phase is close to one million years in this case.

The third intermediate pure proton component produced near SMBH is closely related with the production of
astrophysical neutrinos. This is because it comes from the neutrons generated in $p\gamma $
interactions. The ratio of the total neutron and neutrino energies is close to 5
in this process while the energy
 of individual neutrinos is 25 times smaller than the neutron energy \cite{mucke00}.
This means that the expected metagalactic energy
 flux of neutrinos is 5 times lower than the one of the corresponding protons.
So we can compare the energy flux of
metagalactic component at 25 PeV shown in Figure 3 with the energy
flux $\sim 5\times 10^{-11}$ erg cm$^{-2}$ s$^{-1}$ sr$^{-1}$ of astrophysical neutrinos at PeV energies
\cite{troitsky21}. We found
 that for the model "heavy" the expected flux of the astrophysical neutrinos is
two times  lower than the measured flux.

As for the chemical composition of observed UHECRs the "light" model without a strong enrichment by heavy nuclei
looks more attractive. The enrichment is expected in the  models with an reacceleration of preexisting low energy
cosmic rays in the jet  \cite{caprioli15,kimura18b}. However, it is not easy for background cosmic rays to reach
 the jet itself because the jet flow is surrounded by an extended region of the heated jet gas (cocoon) and heated
at the bow shock interstellar gas (see Figure 1). In such a situation the
reacceleration of particles accelerated at the bow
shock might be more probable \cite{zirakashvili23}.

Future advances in the investigation of the chemical composition of the highest energy
cosmic rays will help to choose the best model. This first concerns the contradictions
between the hadronic interaction models (see Figures 4-6). The dominance of Helium nuclei at the
end of the spectrum will be in favor of the Galactic origin of UHECRs (model "light")
because Helium nuclei with such energies cannot come from the extragalactic sources. The
heavier composition will be in favor of the Galactic model "heavy". An extragalactic origin
of UHECRs is also possible in this case.

Further development of more realistic and less phenomenological
models for particle acceleration
 in jets is also needed since there are deviations of
experimental and simulated $\langle X_{max}\rangle $, $\sigma(X_{max})$ (see Figures 5,6). We leave this
problem for future investigations.

A crucial assumption of our model is the strong magnetic field of microGauss strength in the extended halo.
For lower values of the field it is impossible to explain the end of the observable spectrum at
energies above $10^{18}$ eV. In this case, a contribution at the highest energies from more distant nearby sources
like Cen A radio galaxy \cite{mollerach19} or Andromeda galaxy \cite{zirakashvili23} can play a role.

On the other hand if the extended halo with microgauss magnetic fields indeed exists, then 
the cosmic ray spectrum and chemical composition at energies above $10^{15}$ eV can be explained by 
the recent Eddington-like accretion event in the Galactic center. 
The assumed jet power $10^{44}-10^{45}$ erg s$^{-1}$ is 
 $10^3$ times higher than the one in our Paper I where the Andromeda galaxy makes the 
main contribution to the spectrum of UHECRs. Such a high power is possible if the recent accretion event was 
 similar to the activity of NLSy 1 galaxies (see the discussion above).
This scenario suggested in the present paper seems to be more 
probable because 
 it is clear that some strong energy release in the Galactic center occurred 3-20 million years ago, while 
the time of the ancient SMBH activity in the Andromeda galaxy is unknown. 

The simulated anisotropy is low in the models under consideration. However, this is because
we use the spherical simulation domain and observe cosmic rays close to the center.
Deviations from the spherical symmetry can result in higher anisotropy, especially at the
highest energies.

For example, one can expect such a deviation because of the interaction with the Andromeda
galaxy and because our Galaxy is moving in the direction of Andromeda. SMBH in the
Andromeda galaxy is 50 times more massive than SMBH in the Galactic center. So it is
expected that outflows driven by AGN activity during the growth of Andromeda's SMBH
produced a huge extended halo of the hot gas with a size of several Mpc. The Milky Way's
gaseous halo is smaller in size and located inside a more extended Andromeda's halo (see
Figure 1). In this situation, we expect that the Galaxy is shifted from its gaseous halo center in
the direction of Andromeda. This is because the galactic wind of Andromeda pushed the
Galactic halo during its formation. The motion of the Galaxy in the direction of Andromeda
produced a similar effect \cite{weaver77}. If the magnetic field is lower in the Andromeda's halo, then
the highest energy particles produced in the Galactic center escape easier in the Andromeda
direction. The diffusive flux is directed to Andromeda in this case and we expect to see
anisotropy from the opposite direction which is approximately the direction of the radio
galaxy Cen A. The results for this case are illustrated in the right panel of Figure 7. 
The direction of the anisotropy is from
 the Galactic center at low energies. It changes to the direction opposite to Andromeda 
at high energies. 
Observations of the Auger Collaboration seems to confirm this pattern \cite{aab18,PAO22}.

\section{Conclusion}

Our conclusions are the following:

1) We model the propagation of ultra-high energy particles from the Galactic central source
that was active several million years ago and compare the all-particle spectra, anisotropy, and
chemical composition obtained with observations. If the active source is less than 3 million
years old, the Helium nuclei do not have time for photodisintegration, and a model using
the light source composition ("light" model) is possible. For older sources, severe enrichment
by heavy nuclei is required to explain the observed spectrum of UHECR ("heavy" model).

2) The necessary condition for both models is the effective confinement of particles in the
extended (several hundred kpc in size) Galactic halo with microGauss magnetic fields. It is
expected that this halo was produced by powerful Galactic wind driven by the star formation
and SMBH activity of the young Galaxy. The Galactic magnetic fields were transported to
the halo and amplified by the Cranfill effect (see Appendix).

3) The jet power must be close to the Eddington luminosity in the model "light" to provide a
high enough maximum energy of accelerated particles.  Such a luminosity
is observed at the active phase of jetted NLSy 1 galaxies \cite{d'ammando19}.

4) We expect that the cosmic ray anisotropy at the highest energies depends on the deviation
from spherical symmetry. If this deviation is caused by the interaction with the Andromeda
galaxy, the anisotropy can be expected from the opposite to the Andromeda direction. This is
close to the anisotropy pattern observed by the Pierre Auger Observatory.

\appendix
\section{MHD modeling of the hot magnetized halo}

We performed simplified one dimensional magnetohydrodynamic (MHD)
calculations of the Milky Way halo formation. The effects of
rotation and radiative losses are neglected. MHD equations for the
gas density  $\rho (r,t)$, gas velocity $u(r,t)$, gas pressure
$P_g(r,t)$, and magnetic field $B(r,t)$ in the spherically
symmetrical  case are given by

\begin{equation}
\frac {\partial \rho }{\partial t}+\frac {1}{r^2}\frac {\partial }{\partial r}r^2u\rho =0
\end{equation}

\begin{equation}
\frac {\partial \rho u}{\partial t}+\frac 1{r^2}\frac {\partial }{\partial r}
r^2\left( \rho u^2+P_g+\frac{B^2}{8\pi }\right)
= \frac{2P_g}{r}-g(r)\rho ,
\end{equation}

\begin{equation}
\frac {\partial \varepsilon }{\partial t}+\frac 1{r^2}\frac {\partial }{\partial r}
r^2u\left( \varepsilon +P_g+\frac{B^2}{8\pi }\right)
= -g(r)\rho u,
\end{equation}

\begin{equation}
\frac {\partial B}{\partial t}+\frac 1r\frac {\partial Bur}{\partial r}=0,
\end{equation}
where
$\varepsilon =\frac 12\rho u^2+\frac {P_g}{\gamma _g-1}+\frac{B^2}{8\pi }$
is the total energy density and $\gamma _g=5/3$ is the adiabatic index of the gas. Equations
(A.1-A.3) are the continuity equation, the momentum equation and the
energy equation respectively while the equation (A.4) describes the
evolution of the non-radial components of the magnetic field.

\begin{figure}
\begin{center}
\includegraphics[width=7.0cm]{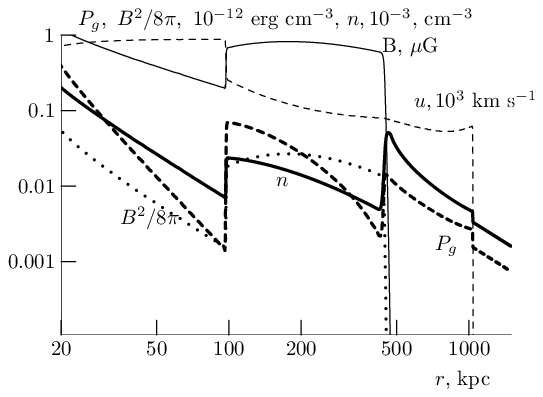}
\hfill
\includegraphics[width=7.0cm]{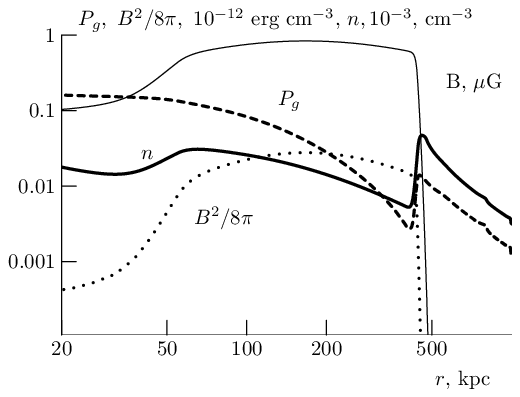}
\end{center}
\caption{ The radial dependence of the gas density (thick solid line), the gas
velocity (thin dashed line),
the magnetic energy density $B^2/8\pi $ (dotted line),
the magnetic field strength $B$ (thin solid line)
and the gas pressure $P_g$ (thick dashed line) at
 $t=5$ Gyr (left panel) and at the current epoch $t=13.7$ Gyr (right panel). }
\end{figure}

The gravitational acceleration $g(r)=V_c^2/r$ is dominated by the dark matter of the virialized
 isothermal halo with parameter $V_c\approx 200$ km s$^{-1}$.
At the initial instant of time $t_0=1$ Gyr after the Big Bang the density and gas pressure are
 given by expressions

\begin{equation}
\rho =\frac {\eta_bV_c^2}{4\pi GR_h^2}
\left\{ \begin{array}{ll}
1, r<R_h\\
R_h^2/r^2, r>R_h,
\end{array} \right.
\end{equation}

\begin{equation}
P_g =\frac {\eta_bV_c^4}{8\pi GR_h^2}
\left\{ \begin{array}{ll}
(1+2\ln (R_h/r)), r<R_h\\
R_h^2/r^2, r>R_h.
\end{array} \right.
\end{equation}
Here $G$ is the gravitational constant and $\eta _b\approx \frac {1}{6}$ is the baryon fraction.
This initial matter distribution corresponds to the situation when the gas in the central part of
 the virialized halo at $r<R_h\approx 150$ kpc was cooled radiatively and formed the Galaxy in the center.
 A half of this mass $\sim 10^{11}M_{\odot }$ will be ejected later leaving the Galaxy with a
baryon deficit ("missing" baryons \cite{mcgaugh10}).

The equations (A.1-A.4) are solved numerically at $r>R_0=15$ kpc. We use the Total Variation Diminishing
hydrodynamic scheme \cite{trac03} with "minmod" flux limiter.
The mass loss rate $25\ M_{\odot }$ yr$^{-1}$ and energy power $8\times 10^{42}$ erg s$^{-1}$ are fixed during 4 Gyr
 at the inner boundary at $r=R_0$. This release of $10^{60}$ erg of energy and $10^{11}M_{\odot }$ of matter
 results in a powerful outflow
(Galactic wind) with the speed of about 900 km s$^{-1}$.
Its magnetization is provided by the magnetic source at the inner boundary.
Its strength is adjusted to obtain the Mach number $M_a=u/V_a=4$ of the wind. The sources are switched off after $t=5$ Gyr.

Figure 8 illustrates the results. The hydrodynamical profiles at the end of the energy release at $t=5$ Gyr
 are shown in the left panel. The magnetic field is compressed at the termination
shock at $r=100$ kpc and is further amplified by the Cranfill effect. As a result,
the magnetic pressure is higher than the gas pressure at the edge of the cavity at $r\sim 500$ kpc.
The expansion of the cavity drives an outer shock at $r\sim 1$ Mpc. At later times the
termination shock goes back to the Galaxy, the reflected shock makes several oscillations and
the system goes to the quasi-steady state at the current epoch, see the right panel. Probably a
weak additional release of energy and matter at $t>5$ Gyr could result in higher values of the
density and magnetic field at distances $r<100$ kpc but can not change the magnetic field and
density distribution at larger distances. We conclude that the microGauss magnetic fields in
the huge Galactic halo are indeed possible.

The final value of the halo magnetic field can be lower for higher values of the Mach number
$M_a$ that is for the lower Galactic wind magnetization. Probably this explains lower values of
the magnetic field strength $B\sim 0.1\mu $G found in 3D MHD cosmological simulations of Milky Way-like
galaxies \cite{pakmor20}. The corresponding simulated rotation measure is
lower than the recently measured Faraday rotation for different samples of galaxies \cite{heesen23,bockmann23}.
In addition, in a real 3-dimensional
geometry the shell of the cavity is unstable relative to the Rayleigh-Taylor instability, and
"fingers" and clouds of the denser outer gas penetrate the cavity.

\acknowledgments We thank an anonymous referee for valuable
comments and suggestions. The work was partly performed at the
Unique scientific installation "Astrophysical Complex of MSU-ISU"
(agreement 13.UNU.21.0007).




\bibliography{jetGC}







\end{document}